\documentclass[onecollarge,natbib]{svjour2}
\bibpunct{[}{]}{;}{n}{}{,} 
\smartqed  
\usepackage{graphicx}
 \usepackage{mathptmx}      

 \usepackage{amsmath}
 \usepackage{amssymb}
 \usepackage{slashed}

\newcommand{\ket}[1]{ | #1 \rangle}
\newcommand{\bra}[1]{ \langle #1 |}

\journalname{Few-Body Systems (EFB22)}

\begin{document}

\title{Meson-cloud effects in the electromagnetic nucleon structure}


\author{Daniel Kupelwieser         \and
        Wolfgang Schweiger
}


\institute{D. Kupelwieser \and W. Schweiger \at
              Institut f\"ur Physik, FB Theoretische Physik, Universit\"at Graz, Austria\\
               \email{daniel.kupelwieser@edu.uni-graz.at}\\
              \email{wolfgang.schweiger@uni-graz.at}           
}

\date{Received: date / Accepted: date}

\maketitle

\begin{abstract}
We study how the electromagnetic structure of the nucleon is influenced by a pion cloud. To this aim we make use of a constituent-quark model with instantaneous confinement and a pion that couples directly to the quarks. To derive the invariant 1-photon-exchange electron-nucleon scattering amplitude we employ a Poincar\'e-invariant coupled-channel formulation which is based on the point-form of relativistic quantum mechanics. We argue that the electromagnetic nucleon current extracted from this amplitude can be reexpressed in terms of pure hadronic degrees of freedom with the quark substructure of the pion and the nucleon being encoded in electromagnetic and strong vertex form factors. These are form factors of bare particles, i.e. eigenstates of the pure confinement problem. First numerical results for (bare) photon-nucleon and pion-nucleon form factors, which are the basic ingredients of the further calculation, are given for a simple 3-quark wave function of the nucleon.
\end{abstract}


\section{Introduction}
\label{intro}
The chiral constituent-quark model, in particular its relativistic version, has been very successful in explaining baryon spectra~\cite{Glozman:1997ag} and the electroweak structure of baryons~\cite{Wagenbrunn:2000es}. Its elementary degrees-of-freedom, which are supposed to emerge as effective particles after spontaneous breaking of chiral symmetry in QCD~\cite{Glozman:1995fu}, are constituent quarks and Goldstone bosons. The latter are realized by  the lightest octet of pseudoscalar mesons. They provide a hyperfine interaction between the constituent quarks on top of a confining linear potential. In Refs.~\cite{Glozman:1997ag,Wagenbrunn:2000es} Goldstone-boson exchange between the quarks has been treated within an instantaneous approximation with the consequence that baryon excitations come out as stable bound states rather than resonances with a finite life time. A  possible way towards overcoming this problem is to take the dynamics of the Goldstone bosons explicitly into account~\cite{
Kleinhappel:2012zj}. This will not only change the baryon mass spectrum, but it will also affect the electroweak structure of the baryons. The physical picture of baryons that emerges from a constituent-quark model with instantaneous confinement and dynamical Goldstone-boson exchange between the constituent quarks is that of a three-quark core which is surrounded by a pseudoscalar meson cloud. This kind of picture is also supported by a recent calculation of electromagnetic nucleon form factors performed within the Bethe-Salpeter-Dyson-Schwinger framework~\cite{Eichmann:2011vu}. The author found \lq\lq clear signals of missing pion-cloud effects\rq\rq\ in the low-$Q^2$ region. In the following we will report on our attempt to estimate the pion-cloud contribution to the nucleon form factors within the framework of a (generalized) constituent-quark model.

\section{Formalism and model }
According to our introductory remarks, we consider a nucleon as a system of three constituent quarks which can emit and absorb pions and interact via an instantaneous confinement potential. \footnote{To make matters not too complicated we neglect other pseudoscalar mesons.}
We make use of the point-form of relativistic quantum mechanics in connection with the Bakamjian-Thomas construction to formulate electron-nucleon scattering in a Poincar\'e-invariant way~\cite{Keister:1991sb}. Following the strategy of Refs.~\cite{Biernat:2011mp,GomezRocha:2012zd,Gomez-Rocha:2013zma} we then calculate the invariant $1\gamma$-exchange amplitude, extract the electromagnetic nucleon current, analyze the covariant structure of the current and identify the electromagnetic nucleon form factors.

In the point-form of relativistic dynamics, all four components of the momentum operator become interaction dependent, whereas the generators of Lorentz transformations stay free of interactions. As a consequence, one has simple rotation and boost properties of wave functions and angular-momentum addition works like in non-relativistic quantum mechanics~\cite{Klink:1998zz}. Thanks to the Bakamjian-Thomas construction, the overall motion of the system can be neatly separated from the internal motion:
\begin{equation}\label{eq:massop}
\hat{P}^{\mu}=\hat{\mathcal M}\,  \hat V^{\mu}_{\mathrm{free}}=
\left(\hat{\mathcal M}_{\mathrm{free}}+ \hat{\mathcal
M}_{\mathrm{int}} \right) \hat V^{\mu}_{\mathrm{free}}\, ,
\end{equation}
i.e. the 4-momentum operator factorizes into an interaction-dependent mass operator and a free 4-velocity operator. Bakamjian-Thomas type mass operators are most conveniently represented in a velocity-state basis.
Velocity states $\vert V;  {\bf k}_1, \mu_1; {\bf k}_2, \mu_2; \dots ; {\bf k}_n, \mu_n\rangle$ are characterized by the overall velocity $V$ ($V_\mu V^\mu=1$) of the system, the CM momenta $\vec{k}_i$ of the individual particles and their spin projections $\mu_i$~\cite{Klink:1998zz}.

Our goal is now to calculate the $1\gamma$-exchange amplitude for elastic electron scattering off a nucleon which is considered as a 3-quark system containing also a 3-quark-pion component. Thereby not only the dynamics of electron and quarks, but also the dynamics of the photon and the pion should be fully taken into account. This is accomplished by means of a multichannel formulation that includes all states which can occur during the scattering process (i.e. $|3q, e \rangle$, $|3q, \pi, e \rangle$, $|3q, e, \gamma \rangle$, $|3q, \pi, e, \gamma \rangle$). What one then needs, in principle, are scattering solutions of
\begin{equation}\label{EVequation}
\left(\begin{array}{cccc}
\hat{M}_{3qe} & \hat{K}_\pi & \hat{K}_\gamma & \hat{K}_{\pi\gamma}
\\
\hat{K}_\pi^\dagger & \hat{M}_{3q \pi e} & \hat{\tilde K}_{\pi\gamma} &
\hat{K}_\gamma \\
\hat{K}_\gamma^\dagger & \hat{\tilde K}_{\pi\gamma}^\dag & \hat{M}_{3q e \gamma} &
\hat{K}_\pi \\
\hat{K}_{\pi\gamma}^\dag & \hat{K}_\gamma^\dagger & \hat{K}_\pi^\dagger &
\hat{M}_{3q \pi e \gamma}
\end{array}\right)
\left(\begin{array}{l}
\ket{\psi_{3q e}} \\ \ket{\psi_{3q \pi e}} \\
\ket{\psi_{3q e \gamma}} \\ \ket{\psi_{3q \pi e \gamma}}
\end{array}\right)
=
\sqrt{s} \left(\begin{array}{l}
\ket{\psi_{3q e}} \\ \ket{\psi_{3q \pi e}} \\
\ket{\psi_{3q e \gamma}} \\ \ket{\psi_{3q \pi e \gamma}}
\end{array}\right)
\end{equation}
which evolve from an asymptotic electron-baryon in-state $\ket{e\,B}$ with invariant mass $\sqrt{s}$. The diagonal entries of this matrix mass operator contain, in addition to the relativistic kinetic energies of the particles in the particular channel, an instantaneous confinement potential between the quarks. The off-diagonal entries are vertex operators which describe the transition between the channels. Making use of the velocity-state representation, these vertex operators can be related to  usual quantum-field theoretical interaction Lagrangean densities~\cite{Klink:2000pp}. The 4-vertices $\hat{K}_{\pi\gamma}^{(\dag)}$ and $\hat{\tilde{K}}_{\pi\gamma}^{(\dag)}$ show up only for pseudovector pion-quark coupling and vanish for pseudoscalar pion-quark coupling. It is now helpful to reduce Eq.~(\ref{EVequation}) to an eigenvalue problem for
$\ket{\psi_{3q e}} $ alone, which can be done by means of a Feshbach reduction:
\begin{equation}\label{eq:Mphys}
\left[\hat{M}_{3qe} +\hat{K}_\pi(\sqrt{s}-\hat{M}_{3q\pi e} )^{-1} \hat{K}_\pi^\dag + \hat{V}_{1\gamma}^{\rm{opt}}(\sqrt{s})\right] \ket{\psi_{3q e}} = \sqrt{s} \, \ket{\psi_{3q e}} \, , \vspace{-0.3cm}
\end{equation}
where
\begin{eqnarray}\label{eq:V1g}
\hat{V}_{1\gamma}^{\rm{opt}}(\sqrt{s})&=& \hat{K}_\gamma\, (\sqrt{s}-\hat{M}_{3qe\gamma}-\hat{K}_\pi (\sqrt{s}-\hat{M}_{3q\pi e\gamma} )^{-1} \hat{K}^\dag_\pi)^{-1} \hat{K}_\gamma^\dag \\
& &+ \hat{K}_\pi (\sqrt{s}-\hat{M}_{3q\pi e})^{-1} \hat{K}_\gamma\, (\sqrt{s}-\hat{M}_{3q\pi e\gamma}-\hat{K}^\dag_\pi(\sqrt{s}-\hat{M}_{3q e\gamma} )^{-1} \hat{K}_\pi)^{-1} \hat{K}_\gamma^\dag (\sqrt{s}-\hat{M}_{3q\pi e})^{-1}  \hat{K}_\pi^\dag+\dots \nonumber
\end{eqnarray}
is the 1$\gamma$-exchange optical potential. The 1$\gamma$-exchange amplitude is now obtained by sandwiching $\hat{V}_{1\gamma}^{\rm{opt}}(\sqrt{s})$ between physical electron-baryon states  $\ket{e\,B}$, i.e. eigenstates of $[  \hat{M}_{3qe} +\hat{K}_\pi(\sqrt{s}-\hat{M}_{3q\pi e} )^{-1} \hat{K}_\pi^\dag ]$. The first term in Eq.~(\ref{eq:V1g}) corresponds to the photon coupling directly to the 3-quark component of the physical baryon, the second term is representative for one of the possible time orderings for the photon coupling to either the pion or the  bare baryon within a pion loop. The crucial point is now to observe that, due to instantaneous confinement, the propagating states do not contain free quarks, they rather contain either physical baryons or bare baryons, the latter being eigenstates of the confinement potential alone. This allows us to reformulate the whole problem as a pure hadronic problem with the quark substructure being hidden in vertex form factors.
\begin{figure*}[!t]
\begin{minipage}{.29\textwidth}
\includegraphics[width=\textwidth]{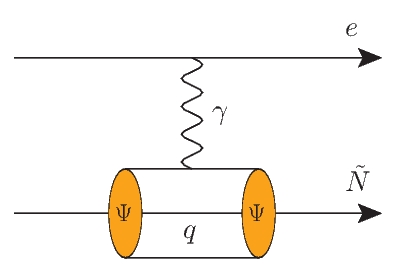}
\end{minipage}
\begin{minipage}{.35\textwidth}
\includegraphics[width=\textwidth]{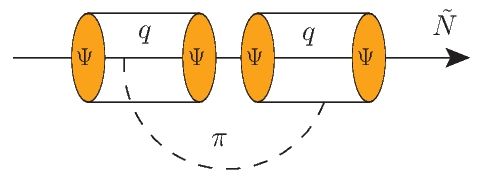}
\end{minipage}
\hspace{1em}
\begin{minipage}{.3\textwidth}
\caption{Quark-level diagrams from which the electromagnetic (left) and strong (right) form factors of the ``bare'' nucleon $\tilde N$ are extracted. The left diagram represents elastic electron scattering off a bare nucleon. The two time orderings of the $\gamma$-exchange are subsumed under a covariant photon propagator. The right diagram is one of the contributions to the kernel of the mass eigenvalue problem one ends up with when coupling a pion to a bare nucleon (cf. Ref.~\cite{Kleinhappel:2012zj}).  $\psi$ is the 3-quark wave function of the bare nucleon.
\label{quarkleveldiags}}
\end{minipage}
\end{figure*}

\begin{figure*}[!t]
\centering
\begin{minipage}{0.25\textwidth}
\vspace{-1.5em}
\includegraphics[width=0.9\textwidth]{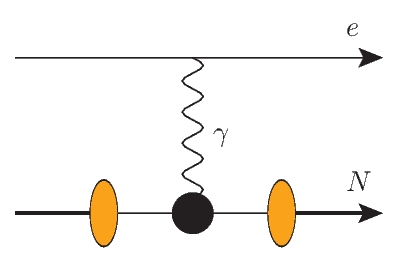}
\end{minipage}
\begin{minipage}{1ex}
\vspace{-1em}
+
\end{minipage}
\begin{minipage}{0.30\textwidth}
\includegraphics[width=0.9\textwidth]{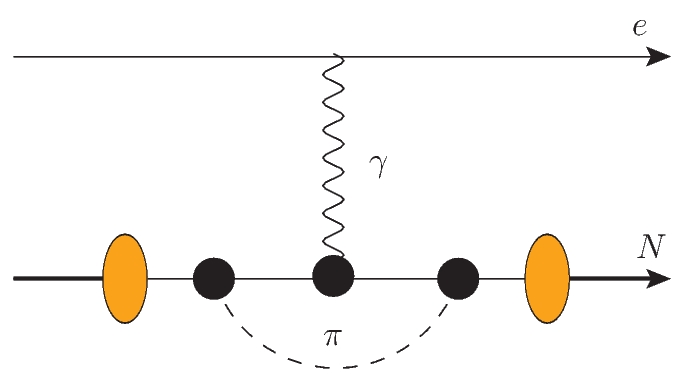}
\end{minipage}
\begin{minipage}{1ex}
\vspace{-1em}
+
\end{minipage}
\begin{minipage}{0.30\textwidth}
\vspace{-1.0em}
\includegraphics[width=0.9\textwidth]{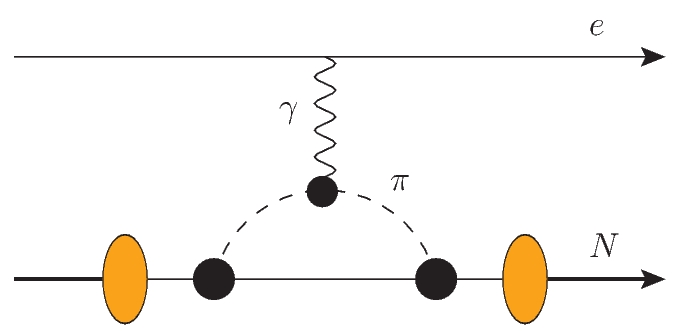}
\end{minipage}
\caption{Diagrams representing the $1\gamma$-exchange amplitude for electron scattering off a \lq\lq physical\rq\rq\ nucleon $N$, i.e. a bare nucleon dressed by a pion cloud ($\gamma\pi\tilde N\tilde N$ vertices not shown). The time orderings of the $\gamma$-exchange are subsumed under a covariant photon propagator. Black blobs represent vertex form factors for the coupling of a photon or pion to the bare nucleon. A vertex form factor is also assumed at the photon-pion vertex. The ovals represent the wave function (i.e. essentially the probability $P_{\tilde N/N}$) for finding the bare nucleon in the physical nucleon. For the correct normalization graphs 2 and 3 have to be multiplied with $(1-P_{\tilde N/N})/P_{\tilde N/N}$.\label{overallff}}
\end{figure*}

If we concentrate on nucleon form factors and neglect nucleonic excitations within the pion loop one just has to replace \lq\lq 3q\rq\rq\ by \lq\lq $\tilde N$\rq\rq\ in Eqs.~(\ref{EVequation}), (\ref{eq:Mphys}) and (\ref{eq:V1g}). $\tilde N$ denotes now a bare nucleon, i.e. an eigenstate of the pure confinement problem (without pionic component). In addition, the vertex operators have to be replaced by the most general vertex one could write down for the coupling of a photon or pion to a bare nucleon. These vertices are a sum of covariants with invariant functions, the form factors, in front. The problem of calculating the electromagnetic current and form factors of a nucleon consisting of 3 constituent quarks and an additional $3q\pi$ component can thus be split into two steps:
\vspace{-2pt}
\begin{itemize}
\item[i)] Calculation of $\gamma\tilde N\tilde N$, $\pi\tilde N\tilde N$, $\gamma\pi\tilde N\tilde N$ and $\gamma\pi\pi$ vertex form factors on the quark level along the lines of Refs.~\cite{Biernat:2011mp,GomezRocha:2012zd,Gomez-Rocha:2013zma}.\footnote{For a discussion of wrong cluster properties associated with the Bakamjian-Thomas construction, the consequences for form-factor calculations and how we deal with them, we also refer to these papers.} $\gamma\tilde N\tilde N$ and $\pi\tilde N\tilde N$ form factors can, e.g., be extracted from the processes depicted in Fig.~1.
\item[ii)] Calculation of appropriate on-shell matrix elements $\bra{e'N'}\hat{V}_{1\gamma}^{\rm{opt}}(\sqrt{s})\ket{e\,N}$ on the hadronic level (see Fig.~2) with the vertex form factors of step i), identification of the electromagnetic nucleon current and extraction of the nucleon form factors.
\end{itemize}

\vspace{-0.5cm}
\section{Numerical results \& outlook}
Until now, we have evaluated the $\gamma\tilde N\tilde N$ and $\pi\tilde N\tilde N$ vertex form factors that are needed as input for the $1\gamma$-exchange amplitude in Fig.~2. For simplicity and later comparison with a corresponding front-form calculation~\cite{Pasquini:2007iz}, we have taken an $SU(6)$ spin-flavor symmetric wave function for the (bare) nucleon with a simple ansatz for the momentum part:
\vspace{-10pt}
\begin{equation}
\label{eq:nucleonwf}
\psi(\vec{\tilde k}_1,\vec{\tilde k}_2,\vec{\tilde k}_3) \propto ((\sum_{k=1}^3\tilde{\omega}_k)^2+\beta^2)^{-\gamma}\, ,
\quad
\tilde{\omega}_k = \sqrt{\tilde{\vec{k}}^2+m_q^2}\quad \hbox{and} \quad \sum_{i=1}^3 \, \vec{\tilde k}_i=0\, .\vspace{-2pt}
\end{equation}
With $m_q=0.263$~GeV, $\beta = 0.607$~GeV and $\gamma=3.5$, a reasonable reproduction of the proton magnetic and electric form factors is already achieved without pionic contribution (see Fig.~3). The results for the strong $\pi\tilde N\tilde N$ form factor depend on whether a pseudoscalar or pseudovector pion-quark coupling is assumed (see Fig.~4). Better agreement with corresponding lattice predictions can be observed with the pseudovector pion-quark coupling which seems to be preferred also due to chiral-symmetry arguments. With these form factors and a model for the electromagnetic pion form factor (e.g. the one from Ref.~\cite{Biernat:2011mp}), we have now the ingredients to calculate the graphs shown in Fig.~2 and will get a first estimate for the pionic contribution to the electromagnetic nucleon form factors. Refinements, like including a $\gamma\pi\tilde N\tilde N$ vertex (with corresponding form factor), taking into account nucleonic excitations within the pion loop, or using a more 
sophisticated 3-quark wave function for $\tilde N$ will then be the subject of further studies.

\begin{figure*}[t!]
\centering
\begin{minipage}{0.56\textwidth}
\includegraphics[width=.8\textwidth]{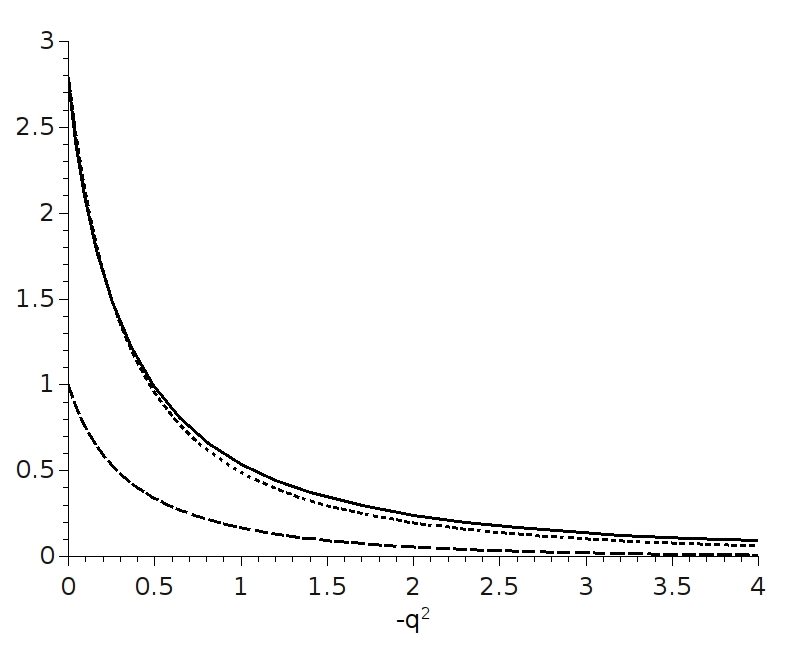}
\caption{Our predictions for the magnetic (solid line) and electric (dashed line) form factors of the bare proton obtained with the simple wave function given in Eq.~(\ref{eq:nucleonwf}). These form factors are renormalized by $P_{\tilde N/N}$ when the pion cloud is included. The dotted lines are phenomenological fits of the corresponding experimental data obtained by Kelly~\cite{Kelly2004}. For the electric form factor the Kelly curve is not visible, because it coincides with our result.\label{elmagfig}}
\end{minipage}
\hspace{1ex}
\begin{minipage}{0.42\textwidth}
\includegraphics[width=.8\textwidth]{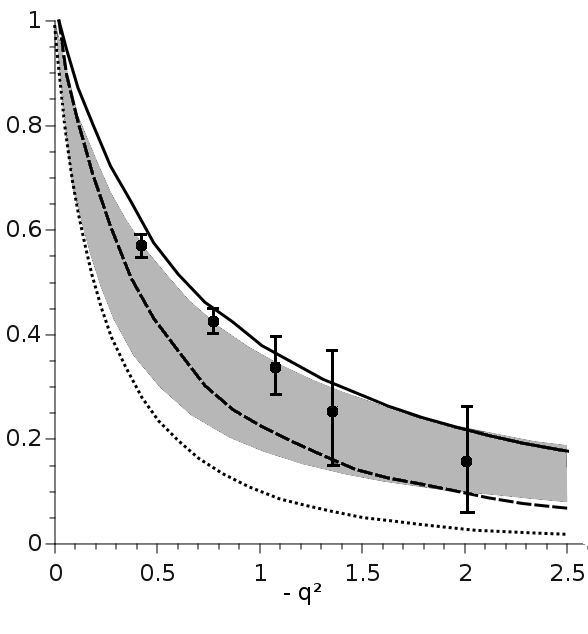}
\caption{Our predictions for the strong $\pi\tilde N\tilde N$ form factor (normalized to $1$ at $Q^2=0$) obtained with the simple nucleon wave function given in Eq.~(\ref{eq:nucleonwf}) and an elementary pion that couples to the quarks by means of either a pseudovector (solid line) or a pseudoscalar (dashed line) vertex. Shown for comparison are another point-form result obtained with a particular ansatz for the pseudoscalar nucleon current and a more sophisticated nucleon wave function~\cite{Plessas2009} (dotted line) and results from lattice QCD by Erkol et al.~\cite{Erkol2009} (grey band) and Liu et al.~\cite{Liu1995} (dots with error bars).
\label{strongfig}}
\end{minipage}
\end{figure*}

\vspace{-0.3cm}
\begin{acknowledgements}
D. Kupelwieser acknowledges the support of the University of Graz and the \lq\lq Fonds zur
F\"orderung der wissenschaftlichen Forschung in \"Osterreich\rq\rq\ (FWF
DK W1203-N16).
\end{acknowledgements}


\end{document}